\documentclass[preprint,aps,superscriptaddress]{revtex4}
\usepackage{graphicx,epsfig,amssymb,amsmath}
\usepackage{placeins}
\usepackage[usenames]{color}
\newcommand*{\field}[1]{\mathbb{#1}}%

\begin{document}
\title{Delay-induced resonance suppresses damping-induced unpredictability}

\author{Mattia Coccolo}

\address{Nonlinear Dynamics, Chaos and Complex Systems Group, Departamento de F\'{i}sica \\ Universidad Rey Juan Carlos, Tulip\'{a}n s/n, 28933 M\'{o}stoles, Madrid, Spain}

\author{Julia Cantis\'{a}n}

\address{Nonlinear Dynamics, Chaos and Complex Systems Group, Departamento de F\'{i}sica \\ Universidad Rey Juan Carlos, Tulip\'{a}n s/n, 28933 M\'{o}stoles, Madrid, Spain}

\author{Jes\'{u}s M. Seoane}

\address{Nonlinear Dynamics, Chaos and Complex Systems Group, Departamento de F\'{i}sica \\ Universidad Rey Juan Carlos, Tulip\'{a}n s/n, 28933 M\'{o}stoles, Madrid, Spain}

\author{S. Rajasekar}
\address{School of Physics, Bharathidasan University, Tiruchirapalli 620024, Tamilnadu, India\\}

\author{Miguel A.F. Sanju\'{a}n}

\address{Nonlinear Dynamics, Chaos and Complex Systems Group,
Departamento de F\'{i}sica \\ Universidad Rey Juan Carlos, Tulip\'{a}n s/n, 28933 M\'{o}stoles, Madrid, Spain}

\date{\today}

\begin{abstract}
Combined effects of the damping and forcing in the underdamped time-delayed Duffing oscillator are considered in this paper. We analyze the generation of a certain damping-induced unpredictability, due to the gradual suppression of interwell oscillations. We find the minimal amount of the forcing amplitude and the right forcing frequency to revert the effect of the dissipation, so that the interwell oscillations can be restored, for different time delay values. This is achieved by using the delay-induced resonance, in which the time delay replaces one of the two periodic forcings present in the vibrational resonance. A discussion in terms of the time delay of the critical values of the forcing for which the delay-induced resonance can tame the dissipation effect is finally carried out.
\end{abstract}

\maketitle

\section{Introduction}\label{sec:1}

The effects of the linear dissipation on both linear and nonlinear oscillators are well known\cite{Kovacic, Johannessen, Trueba}. In this sense, the amplitude of the oscillations decays and eventually goes to zero more or less rapidly depending on the magnitude of the damping term. However, when the dissipation competes with an external forcing, oscillations may survive or decay, depending on the intensities of both terms. In fact, this competition gives birth to different effects on the dynamics of an oscillator, as studied in Ref.~\cite{Mattia}, where it has been analyzed how the dissipation can introduce uncertainty in the topology of the phase space of the
system and how the forcing can counter it. Here, we aim to extend
the current knowledge on the effects of dissipation for time-delayed
oscillators. This kind of systems include a term that depends on a
time interval $ [-\tau, 0] $ of the history of the system, where
$\tau$ is the time delay. This term may destroy stabilities
\cite{Gumowski} and induces oscillations on the system depending on
its parameters. In physical and biological systems, the time delay
accounts for the finite propagation time as information is not
immediately propagated in nature. This means that the future
evolution of the system depends not only on its present state, but
on its previous states. This is why we talk about history functions
instead of initial conditions. History functions are sets of initial
conditions in the continuous time interval $[-\tau, 0]$ (see
Refs.~\cite{Farmer1982}). For simplicity, in this paper, we use as
history function: $ u_{0}=1 $. Time-delayed systems can be found in
many practical problems. Among others, they are present in neural
networks \cite{Popovych}, population dynamics
\cite{Kuang1993a,Liu2016},  electronics \cite{Just2010a} or
meteorology \cite{Keane2017}.

We use the underdamped time-delayed Duffing oscillator due to the paradigmatic role played in nonlinear dynamics. In the first part of this work, we explore the unpredictability induced by the dissipation and relate it with different values of the time delay. Thus, we show that the uncertainty can be reverted when the phenomenon of resonance is triggered. The phenomenon of resonance in nonlinear systems has been deeply studied, for instance, in Ref.~\cite{RaSan}.
In particular, vibrational resonance (VR) \cite{VR, RaSan2} is a well-known resonance phenomenon in which the amplitude enhancement is triggered by two periodic external forcings with different
frequencies. The delay term can play the role of one of the periodic forcings typically used in vibrational resonance, and the resonance can be triggered by the cooperation of the time delay and just one external periodic forcing as shown in Ref.~\cite{Lv}. The latter phenomenon has been called delay-induced resonance and has been studied among other fields in the context of meteorology, as for example the ENSO model in Ref.~\cite{Julia}. Other branches of science where time delay is relevant (neural networks, population dynamics or electronics, as cited before) will be an important focus of study in the following years.

Here, we start by countering the effects of the dissipation through the introduction of a periodic forcing as small as possible in our time-delayed oscillator. In particular, we show that a certain value of the forcing frequency is the more suitable to induce the resonance.  Then, we continue analyzing the possibility of using the forcing to enhance the single-well oscillations, related with particular values of the time delay $\tau$, so that they are no longer
confined in one of the wells. In this case, we find that there is a critical value of the forcing amplitude that shifts the system sensitivity towards another value of the forcing frequency.

For visualization purposes, we plot color gradients in the parameter $(p_i,p_j)$ space, where $p$ means a generic parameter, each color related to the peak to peak amplitude of the oscillations so that every couple of $(p_i,p_j)$, $i\neq j$ with $i,j\in\field{N}$ is colored with the asymptotic amplitude generated. We call this kind of plot {\it amplitude basins}. With the use of those amplitude basins we show how the damping-induced unpredictability is connected with the suppression of the interwell oscillations due to the appearance of fractal regions in the amplitude basins. Moreover, they permit to easily visualize when the unpredictability is reverted or the single-well oscillations disappear. Also, the analysis of those kind of plots makes it easy to determine the critical values, commented above. Every numerical integration in this paper has been performed using the method of steps to reduce our DDE to a sequence of ODEs which are solved by a Runge-Kutta algorithm with adaptative stepsize control
(Bogacki–-Shampine $3/2$ method).

This paper is organized as follows. We describe the model in Sec.~\ref{sec:2} showing the role of the damping term. The phenomenon of high amplitude oscillations in the two wells with a critical value of the frequency  is shown in Sec.~\ref{sec:3}. On the other hand, we show the phenomenon of delay-induced resonance for the single-well oscillations in Sec.~\ref{sec:4}. Finally, the main conclusions of the present work are presented in Sec.~\ref{sec:5}.

\section{The model and the damping effect}\label{sec:2}

For our analysis, we consider an underdamped Duffing oscillator with a restoring force $\alpha x+ \beta x^3$,  a damping term $ \mu\dot{x} $, a periodic forcing $ F\cos{\Omega t} $, and a time delay term $ \gamma x(t-\tau) $. Thus, the equation reads as follows
\begin{equation}\label{eq:1}
  \ddot{x}+\mu\dot{x}+ \alpha x+ \beta x^3+\gamma x(t-\tau)=F\cos{\Omega t}.
\end{equation}
For convenience, we fix the parameters as $ \alpha=-1 $, $\beta=0.1 $ and  $\gamma=-0.3$ as in Ref.~\cite{RaSan}, so that the system is bistable. As a consequence, the system possesses three equilibria: one unstable point at the origin and two stable points at $x^{*} \approx \pm 3.606 $, which are located at the bottom of the wells. To calculate the equilibria, we set $x(t)=x(t-\tau)=x^{*}$, and we obtain the following results
\begin{equation}\label{eq:3}
  x^{*}=0, x^{*}_{\pm}=\pm\sqrt{\frac{-\alpha-\gamma}{\beta}}=\pm\sqrt{13}=\pm3.606.
\end{equation}

\begin{figure}[t]
\centering
\includegraphics[width=13.5cm,clip=true]{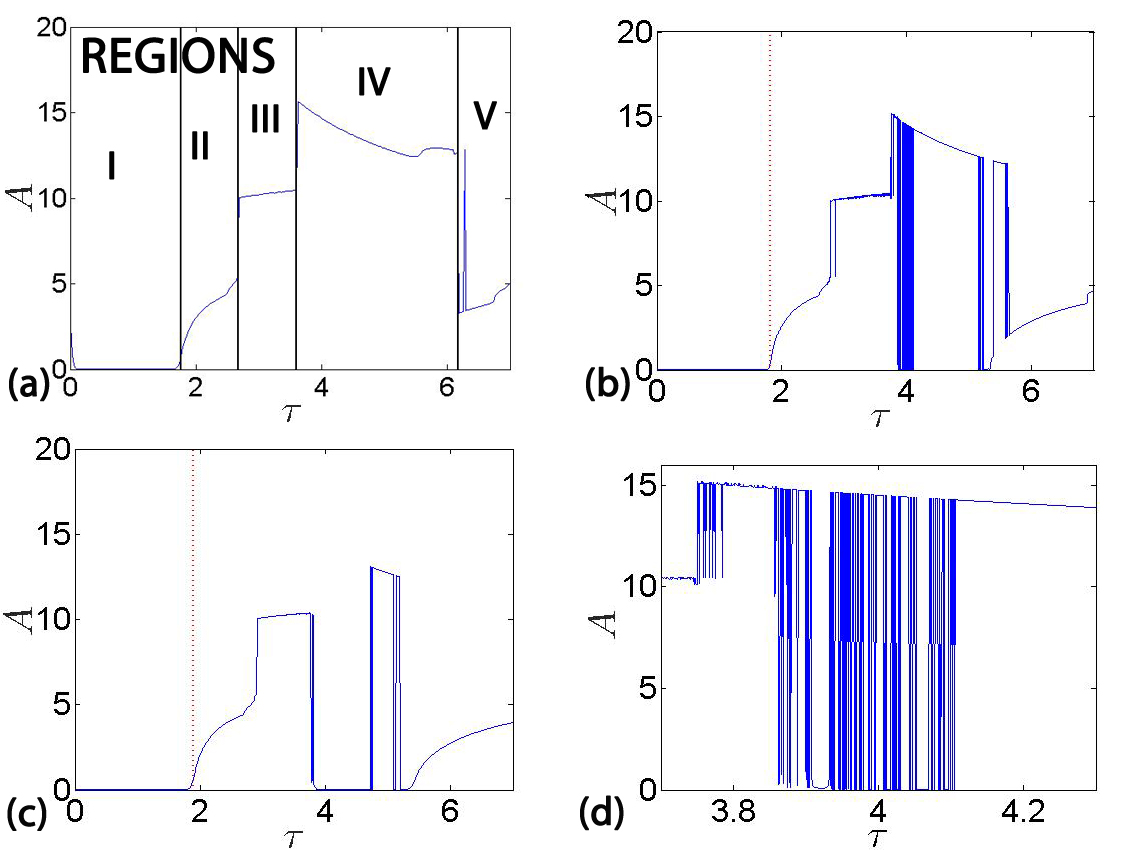}
\caption{The maximum peak to peak amplitude $A$ versus the delay term $\tau$
for equation~\ref{eq:1} with $ F=0 $. (a) The figure shows the variation of the amplitude versus $\tau$ for $\mu =0$, showing the $5$ regions for the different patterns of behaviour of the oscillations amplitude. Panels (b) and (c) plot the same for $\mu=0.02$  and $\mu=0.04$, respectively. Panel (d) is a zoom of panel (b) to better visualize the first $\tau$ values for which there are fluctuations in the trajectories amplitude in the Region IV. In panels (b)-(c) the vertical dotted red lines are the values of $\tau$ predicted by the stability analysis  at which the fixed point $x^*=\pm 3.606$ undergo change of stability.  }
    \label{fig:1}
\end{figure}

Before exploring the combined effects of the forcing, the time delay and the dissipation, we proceed to analyze the dynamics of the oscillator without forcing and delay term. This means that assuming no forcing, $F=0$, and assuming no delay term, $\gamma=0$,  so that the oscillations are damped and confined to one single well due to the dissipation. As previously mentioned, the time delay induces sustained oscillations in the system for certain values of the parameters $ (\gamma, \tau) $. The dependence of the amplitude of these oscillations on $ \tau $ has been studied in the case of absence of dissipation, $ \mu=0 $, in Ref.~\cite{Julia}. The results are summarized in figure~\ref{fig:1}a, where we show the peak to peak values of the amplitude $A$ versus the time delay $\tau$.
%
%
The figure shows several regions where the amplitude behaves differently for different intervals of time delay values. As it can be seen, there is a first region, \textit{Region I}, $\tau\in(0, 1.76)$, with zero amplitude. In this case, the delay does not induce oscillations. When $ \tau $ is increased, oscillations are created but confined to one well. This happens for $\tau\in[1.76, 2.68)$ and we call this interval as \textit{Region II}. In \textit{Region III}, where $\tau\in[2.68, 3.6]$, the trajectories go from one well to another well leading to cross-well motion and so the amplitudes jump to a value bigger than the width of the well. In this region, the motion is aperiodic. We define \textit{Region IV} located at the interval $\tau\in(3.6,6.18)$, where the motion is periodic and trajectories comprise both wells. In the \textit{Region V} indicated in figure \ref{fig:1}a the motion becomes again confined to one well.

\begin{figure}[htp]
    \centering
    \includegraphics[width=11.0cm,clip=true]{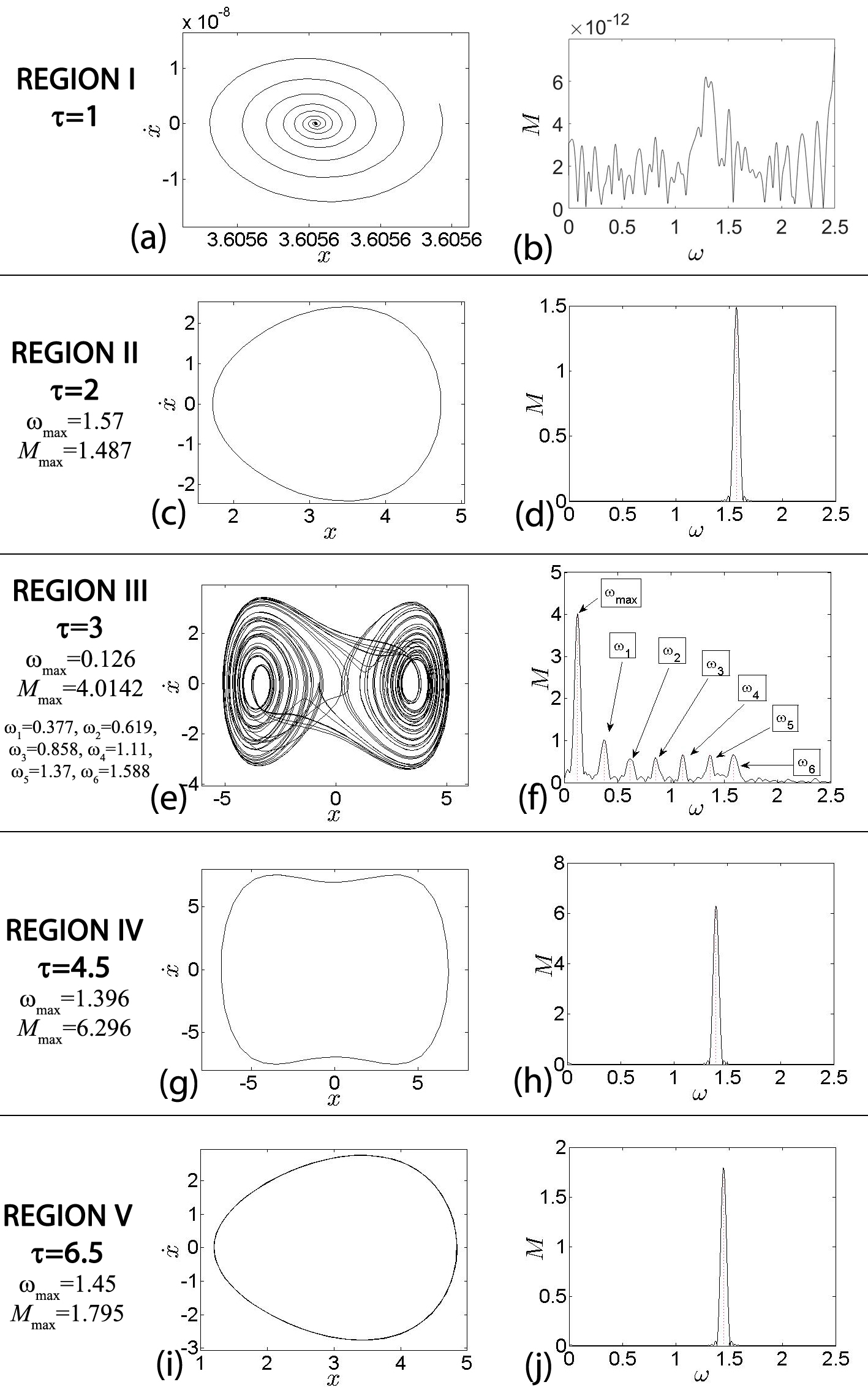}
\caption{The representation of the phase space orbits and frequency spectra
belonging to the different regions shown in figure~\ref{fig:1}a. The computations are done with equation~(\ref{eq:1}), for the parameter values $F=0$ and $\mu=0$. From left to right we represent the orbit in phase space and the frequency spectrum of the oscillations. In panels (a) and (b), we take $\tau=1$, and the asymptotic solution falls into a fixed point and the frequency spectrum shows no oscillations. In panels (c)-(d), $\tau=2$, and the solutions are periodic and confined in one well. In panels (e)-(f), $\tau=3$, and the solutions are aperiodic. In panels (g)-(h), $\tau=4.5$, and the solutions are sustained interwell periodic orbits. Finally, in panels (i)-(j), $\tau=6.5$, and the solutions are again confined in one well.
In the legend the maximum amplitude, $M_{\mathrm{max}}$, and the trajectories related frequencies,  $\omega_{\mathrm{max}}$, are displayed. }
    \label{fig:2}
\end{figure}

For clarity, the dynamics of the system in all the above regions, when $ \mu=0 $, is displayed in figures~\ref{fig:2}a-j.
%
%
\begin{figure}[htp]
    \centering
    \includegraphics[width=7cm,clip=true]{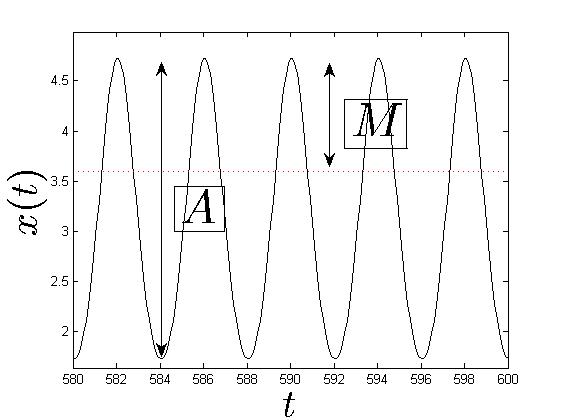}
\caption{The difference between the peak to peak amplitude $A$ and the amplitude  $M$ shown in the FFT spectrum. The dotted line shows the fixed point $x^*=3.606$. }
    \label{fig:2bis}
\end{figure}
%
On the left column, we display the orbits in the phase space, while on the right column we display the frequency spectra calculated through the Fast Fourier Transform (FFT), for each of the regions. Naturally, for the first region, for which the oscillations decay with time to one of the two fixed points $x^*_{\pm}$, the FFT calculation of the frequency shows no peaks, i.e., no periodicity. For the rest of the regions, it is important to notice that the amplitude does not correspond to the peak to peak amplitude $A$ as before. On the contrary, this amplitude $M$ is calculated as $x_{\mathrm{M}}-x^*$, where $x_{\mathrm{M}}$ is the maximum amplitude value and $x^*$ is the stable fixed point. To better show the difference between the amplitudes $A$ and $M$ we have depicted  figure~\ref{fig:2bis}. In the legend of the panels, we show the values of the maximum amplitude, $M_{\mathrm{max}}$, of the peak and the corresponding frequency, $\omega_{\mathrm{max}}$, which is the frequency of the delay-induced oscillations.

Now, we want to go further and explore the effect of the time delay on the dynamics when the damping $ \mu \neq 0 $. In the figures~\ref{fig:1}b-c, we can observe the amplitude variation with $ \tau $ for the damping parameter values $\mu=0.02$ and $\mu=0.04$. As in the case $\mu=0$ described above, a similar pattern of regions for the amplitude versus $\tau$ is observed, though for higher values of $\mu$ the dynamics changes. As a matter of fact, we can see how the high-amplitude oscillations, corresponding to Region IV, either disappear and the system remains at rest at a low energy as in the case  $\mu=0.04$ or it happens for a few $\tau$ values as in the case  $\mu=0.02$. To stress out this damping effect, we show in figure~\ref{fig:1}d a zoom of the Region IV for the $\mu=0.02$ case.

In order to provide analytical support for the numerical results presented above, we perform a linear stability analysis \cite{Lakshmanan,hamdi:2012} for the fixed points $x^{*}=x^*_{\pm}=\pm3.606$. Even though the analysis is performed for $x^{*}=+3.606$, for symmetry reasons it is equivalent for $x^{*}=-3.606$. The characteristic equation of the linearized system reads
\begin{equation}\label{eq:4}
  \lambda^2+\mu\lambda+\alpha+3\beta (x^{*})^2+\gamma \mathrm{e}^{\lambda\tau}=0.
\end{equation}
We take $\lambda = \rho + \mathrm{i}\omega$  as the eigenvalue associated with the equilibria $x^{*}_{\pm}$. The critical stability curve is the one for which $\rho = 0$ as it implies a  change of sign in  $\mathrm{Re}(\lambda)$. For $\mathrm{Re}(\lambda)$ < 0 , the fixed point is stable while for $\mathrm{Re}(\lambda)$ >0  is unstable. Substituting $\lambda = \mathrm{i}\omega$ in equation~\ref{eq:4} and separating both the real and imaginary parts, we obtain the following equations
\begin{align}
\label{eq:5} \omega^2-\alpha -3 \beta (x^{*})^2 &= \gamma \cos{\omega \tau} \\
 \label{eq:6} \mu  \omega &= \gamma  \sin{\omega \tau}.
\end{align}
After squaring and adding both equations, and substituting the parameter values $\alpha= -1$, $\beta=0.1$, $\gamma=-0.3$, and  $x^{*} \approx \pm 3.606 $, we obtain an equation relating the positive values of $ \omega $ in terms of the damping parameter $\mu$ given by
\begin{align}
  \label{eq:7} \omega_{1,2}&=\sqrt{ \frac{-5 \mu^{2} \mp \sqrt{25\mu^{4}-290\mu^{2}+9}+29} {10}},
\end{align}
where $\omega_1$ and $\omega_2$ are the two positive values. We can easily derive
\begin{equation}\label{eq:8}
\tau=\frac{\arccos(\frac{\omega^{2}-\alpha-3 \beta
(x^{*})^{2}}{\gamma})}{\omega}
\end{equation}
from  equation~\ref{eq:5}, providing the $\tau$ values for which the stability changes in terms of the parameters and the frequency. Note that due to the periodicity of $\arccos$ function, we can consider as solutions either $\tau$ or $\frac{2\pi}{\omega} - \tau$. From equation~\ref{eq:7}, we can obtain the values $\omega_{1} = 1.613$ and $\omega_{2} = 1.788$ when $\mu=0.02$, and $\omega_{1} = 1.615$ and $\omega_{2} = 1.786$ when $\mu=0.04$. The minimal value of $\tau$ in the range considered that is obtained from these frequencies is $1.8211$ for $\mu=0.02$ and $1.8943$ for $\mu=0.04$. These values match with the numerically calculated values of $\tau$ for which the fixed points lose stability as shown by the vertical dotted red lines in figures~\ref{fig:1}b-c. Besides, they mark the end of Region I. Notice that the stability also changes for higher values of $ \tau $.

%
\begin{figure}[htp]
\centering
   \includegraphics[width=13.5cm,clip=true]{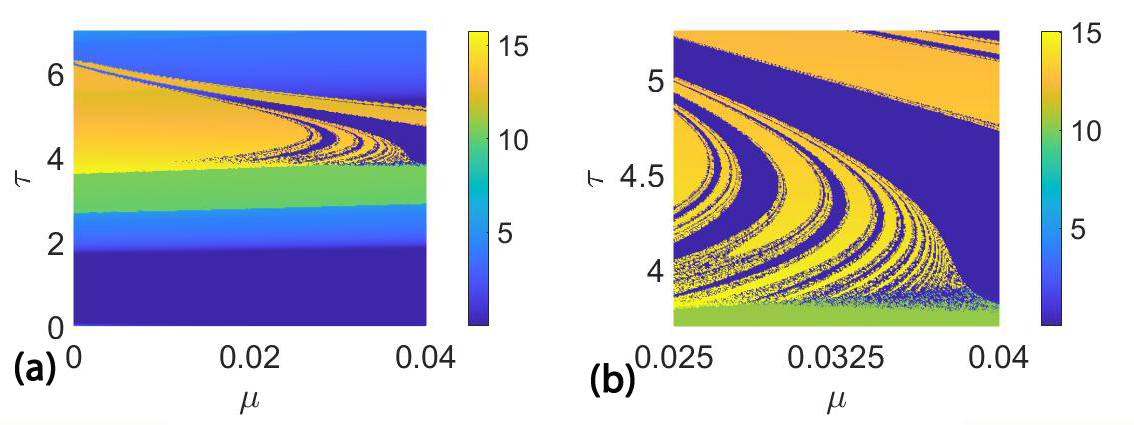}
\caption{Amplitude basins: the figure shows a color gradient plot of the peak to peak amplitude of the oscillations in the parameter space $(\mu,\tau)$ for the case when the forcing amplitude is $F=0$. Panel (b) is a zoom corresponding to Region IV, where $0.025<\mu<0.04$ and $3.6<\tau<5.25$. It is interesting to notice how the fractalization of the yellow basin becomes more and more important as the dissipation term grows.}
\label{fig:3}
\end{figure}
%

We analyze in detail the effect of the damping on the time-delayed system when $ F=0 $ in figure~\ref{fig:3}.  The amplitude is plotted for different $(\mu,\tau)$ values; from now on we refer to this kind of plot as {\it amplitude basins} and the amplitude of the oscillations are calculated peak to peak for an easier display of the amplitude basins. For particular values of $ \mu $, vertical slices, we recover the same pictures as in figure~\ref{fig:1}. In this case, from the bottom to the top, the dark blue is the Region I without oscillations, in the light blue area live the single-well oscillations (Region II), the green area corresponds to interwell aperiodic oscillations (Region III) and the yellow one to high-amplitude and periodic interwell oscillations (Region IV). It is remarkable how the damping only appears to affect the latter region transforming it into the dark blue region where the amplitude is zero. Finally, we have Region V, that also remains untouched by the damping.

Figure~\ref{fig:3}b shows a zoom of figure~\ref{fig:3}a, for values of the time delay and dissipation that generate a more complex structure in the parameter space. Regions I and IV do not present smooth basin boundaries like Regions I and II, for instance. On the contrary, it can be seen in this zoom that the yellow and blue basins are intermingled leading to a non-integer fractal dimension for the basin boundary. So that, a little variation in the parameters $ (\mu,\tau) $ can change the motion of the system from
high-amplitude oscillations to staying at rest, without passing by intermediate states. In other words, the damping produces a fractalization of the amplitude basins, which implies a higher
unpredictability for the system, showing only two possible states, interwell oscillations or no oscillations at all. We refer to this phenomenon as damping-induced fractalization.

\section{Restoring the interwell high-amplitude oscillation with a minimum value of the forcing parameter $F$}\label{sec:3}

In this section we address the following question: is it possible to restore the high-amplitude oscillations suppressed by the damping effect? The previous section showed how the damping in our system (equation~\ref{eq:1} with $ F=0 $), reduces the high-amplitude oscillations of Region IV (yellow in figure~\ref{fig:3}) making the trajectories to fall into the fixed point (dark blue in figure~\ref{fig:3}). This change may be undesired, specially because a small variation in the time-delay may cause a dramatic change in the dynamics. In fact, in the amplitude basins the yellow and blue basins are intermingled due to the damping-induced fractalization of the parameter space.

The phenomenon of delay-induced resonance studied in \cite{Julia} for the overdamped case implies that even for the parameter values for which the time delay does not induce sustained oscillations, a resonance may appear following a different mechanism. Our scenario is different, as the underdamped oscillator presents oscillations and the damping term, for certain values, eliminates these oscillations. However, we show that even in this case, the delay-induced resonance phenomenon appears and as a consequence a small forcing is a valid mechanism to gain back the oscillations and reduce the fractality caused by the damping.

In the following subsections, we explore the parameter values for which a small periodic forcing can restore the high-amplitude oscillations. To achieve our goal, we start analyzing, in the parameter space $(\Omega,\tau)$,  the interaction between the damping parameter $\mu$ and the forcing amplitude $F$. Then, for fixed $\mu$ and $\tau$ values, we study the $(\Omega, F)$ parameter space to evaluate their effect on the oscillations amplitude. Here, we are mainly interested in the Region IV for $ \mu \in [0.025, 0.04] $, for which the oscillations disappear and the fractalization is pronounced. Results are presented in different subsections depending on which parameters are swept.

\subsection{Effect of the frequency of the forcing}

\begin{figure}[htp]
   \centering
   \includegraphics[width=13.5cm,clip=true]{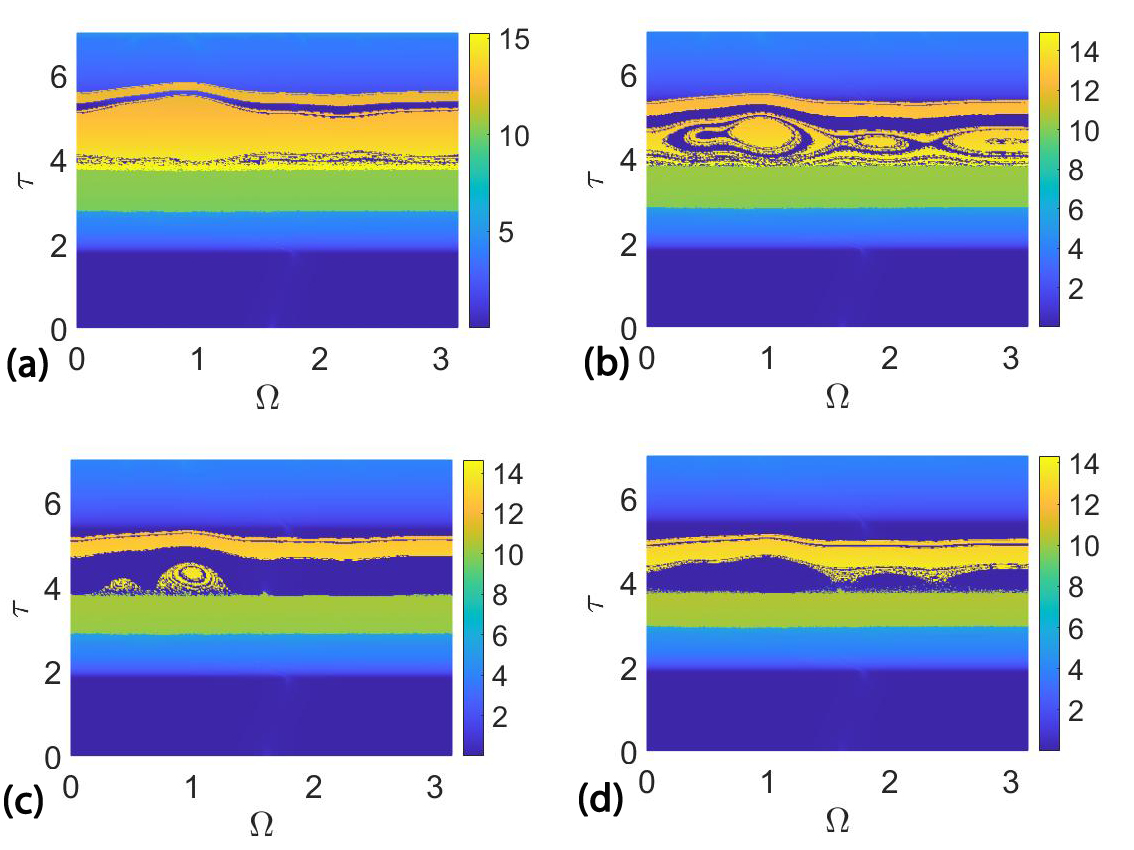}
\caption{The erosion of the yellow amplitude basin in the parameter space $(\tau,\Omega)$ are represented in presence of a forcing of amplitude $F=0.02$. The color bar relates the color with the oscillations amplitude. The panels (a-d) show the basins for $\mu=0.02$, $0.03$, $0.04$ and  $0.05$, respectively. }
\label{fig:4}
\end{figure}

As introduced before, we start analyzing the $(\Omega,\tau)$ parameter space to study the impact of a small periodic forcing amplitude, $F=0.02,$ for fixed damping parameter values. In fact, in figure~\ref{fig:4} the forcing frequency $\Omega$ is varied for four values of $\mu$.
%
%
%
Remember that for $ \mu=0.02 $, when $F=0$, the yellow amplitude basin is still present and the fractalization of the basins only begins at its boundary. On the other hand, for $\mu=0.05$ most of the yellow basin has disappeared. Therefore, we can see that by adding the forcing for the first case, the dynamics does not barely change (figure~\ref{fig:4}a), except for a widening at around $\Omega=1$. On the other hand,  by taking a look to figure~\ref{fig:3} we can see that for $\mu=0.04$ the yellow basins is completely gone. In fact, figure~\ref{fig:4} shows us, as we increase the values of $\mu$, the erosion of the higher amplitude basin. But, it also shows that the effect of introducing a forcing is to create a yellow `island' around $\Omega=1 $. This means that near a specific frequency the forcing is able to counter the damping effects and to recover the high-amplitude interwell oscillations. In figure~\ref{fig:4}c, where $ \mu=0.04$, we can see that still the small high-amplitude island resists and it is centered around that frequency value. So, we call it the resonance frequency $\Omega_{\mathrm{r}}=1$. It is interesting to note that this frequency value is different from the $\omega_{\mathrm{max}}$ observed in figure~\ref{fig:2}h. Now the yellow island disappears for $\mu=0.05$, figure~\ref{fig:4}d, suggesting a shift for the minimal value for which the interwell oscillations disappear when the forcing is present.

Additionally, it is remarkable that the yellow amplitude basin ($\tau \in [3.8, 4.5] $) is the only one that seems to be affected by the presence of the forcing or the damping. This region is less
stable and more susceptible to changes in the dynamics. Oscillations in one single well (light blue basin), the rest state (dark blue basin) or aperiodic interwell oscillations (green basin), at these small values of the forcing or damping, are not modified.

\subsection{Effect of the amplitude of the forcing}

\begin{figure}[htp]
    \centering
    \includegraphics[width=13.5cm,clip=true]{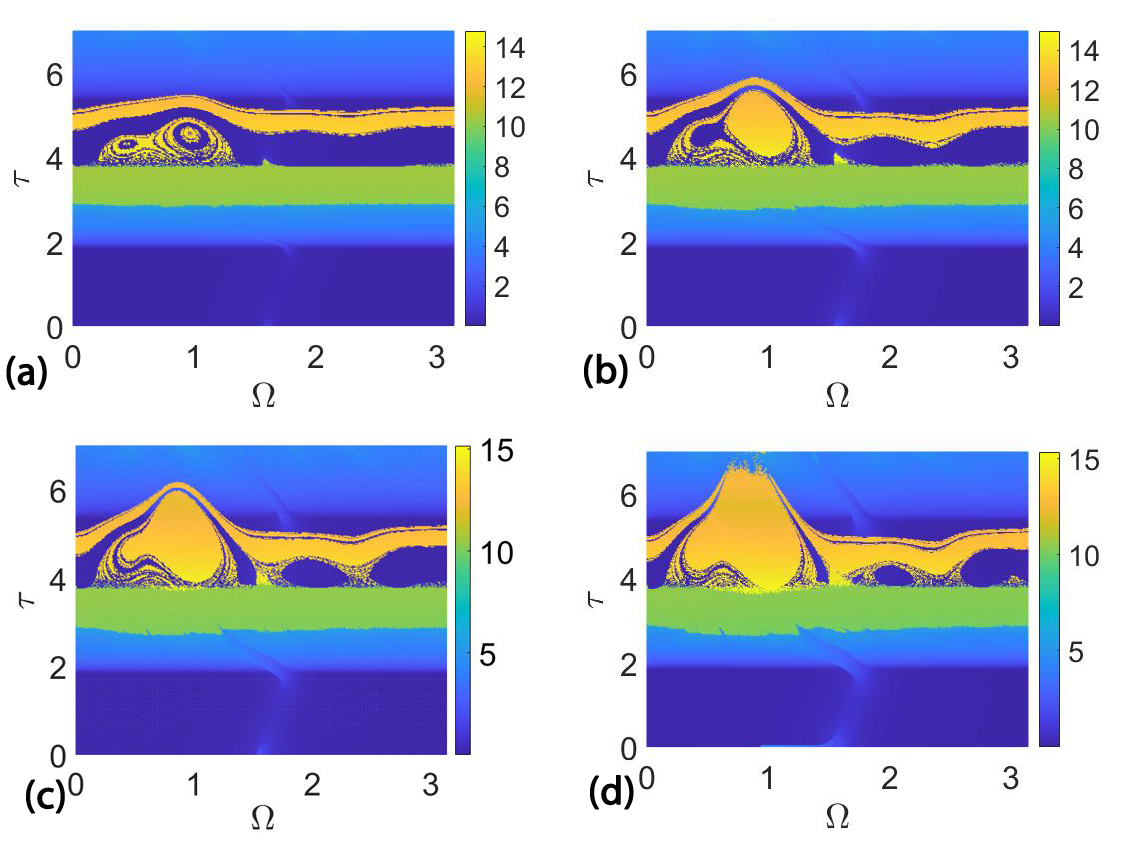}
\caption{The reconstruction of the amplitude basins in the parameter space $(\tau,\Omega)$ are shown for $\mu=0.04$ and in presence of a forcing with   $F=0.04$ (a), $F=0.08$ (b), $F=0.1$ (c) and  $F=0.14$ (d). }
    \label{fig:5}
\end{figure}

In the previous subsection, we have studied how the enhancement of the dissipation, with a fixed small amplitude forcing, suppresses the high-amplitude oscillation except around an area centered in
$\Omega_r=1$. In particular, for the value of $\mu=0.04$, that makes the interwell oscillations to disappear for $F=0$, so that only a little island of high-amplitude oscillations remains. This is the most interesting
case to study, that is, whether the interwell oscillations can be restored with the right forcing amplitude. Therefore, we fix $\mu=0.04$ and plot, in figure~\ref{fig:5}, the amplitude basins in the parameter space $(\Omega,\tau)$ but changing the forcing amplitude.
%
%
%
In these last figures, it is possible to see that the value of $F$, again, only affects the yellow amplitude basin. For the values of $F$ considered ($ F=0.04, 0.08, 0.1, 0.14 $), we would not notice almost any effect if the time delay of our system is outside the Region IV. It is interesting to observe that in figure~\ref{fig:5}d the yellow basins break through the barrier between regions, introducing interwell oscillations in Region V for forcing frequencies near $\Omega\approx1$.

On the contrary, for increasing values of the forcing amplitude, the yellow basin `island' grows up, around the frequency $\Omega_{\mathrm{r}}$, reducing the fractalization. That is, near the resonance frequency the damping-induced unpredictability can be suppressed. On the other hand, for different frequency values even though it is possible to appreciate a reconstruction of the yellow basin, the structure is complex and intermingled. This gives us the relevance of the frequency selection to trigger the delay-induced resonance and to restore the interwell oscillations. It is also worth mention that in the Region IV, there are values of $\tau$ for which, independently of the forcing parameters, the oscillations reach the two wells. We refer to the yellow stripe around $ \tau\approx5$.

\subsection{Effect of the forcing amplitude and frequency for fixed $\mu$ and $\tau$ values}

\begin{figure}[htp]
    \centering
    \includegraphics[width=13.5cm,clip=true]{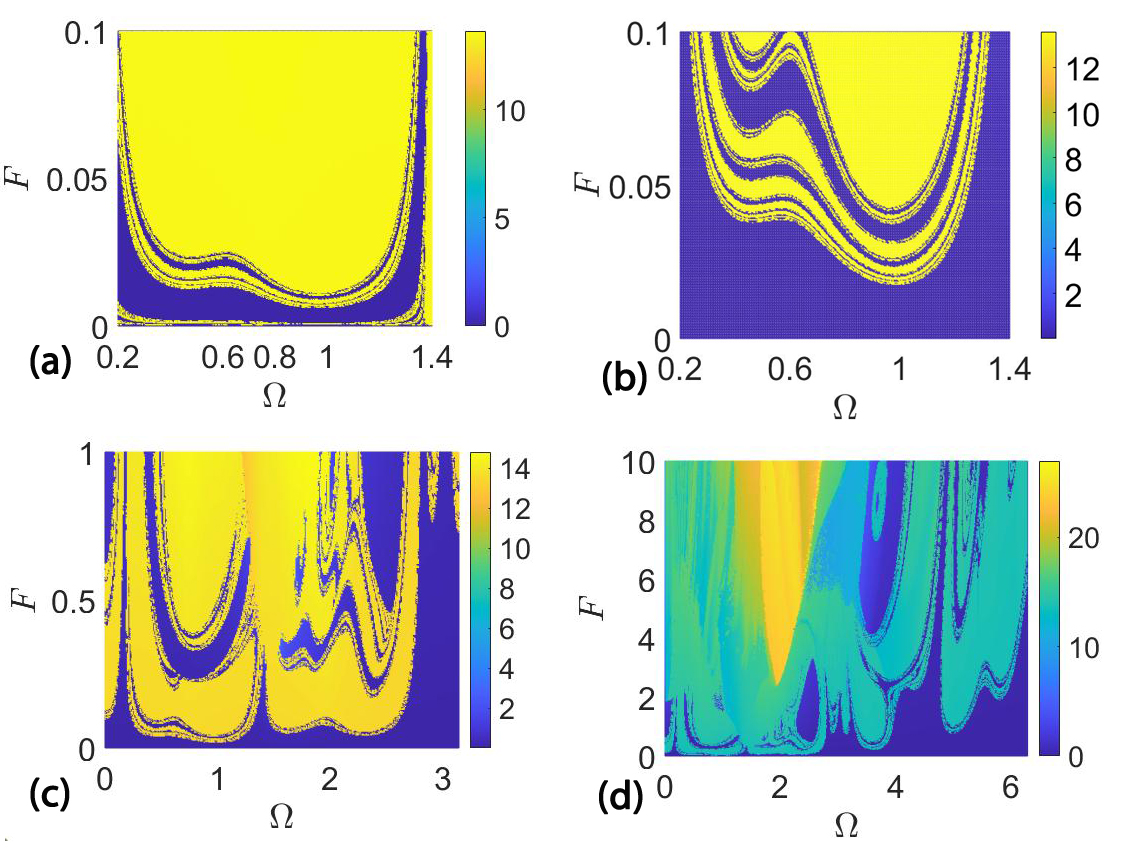}
\caption{Amplitude basins in the parameter space $(\Omega,F)$ for $\tau=4.5$ and for (a) $\mu=0.03$ and (b) $\mu=0.04$. The interval of frequencies has been chosen to enclose the yellow `island' observed in figure~\ref{fig:4}. Panels (c) and (d) are expansions of the panel (b) for further values of the maximum amplitude $F$ and for a wider band of frequencies.}
    \label{fig:6}
\end{figure}

At this point we know the $\mu$ and $\tau$ values where we focus our attention and we aim to explore the parameter space $(\Omega,F)$. Therefore, we fix the value $\tau=4.5$, which is in the
middle of the yellow islands that has been created thanks to the damping forcing competition. In figures~\ref{fig:6}a-b we start focusing our attention around the frequency values that generated
the yellow `island' described before, so $\Omega$ changes in the range of values $[0.2,1.4]$ while $F$ in the range $[0,0.1]$, with $\mu=0.03$ and $\mu=0.04$, respectively.
%
%
%
For $ \mu=0.03 $, we do not need a high forcing, that is $ F \gtrsim 0.03 $, to restore the high-amplitude oscillations for almost any value of $ \Omega $, in the interval kept in consideration. However, for $ \mu=0.04 $, when the forcing is not present, the high-amplitude oscillations disappear completely for the range of $ \tau $ considered and they are only restored for a certain range of values of $ \Omega \gtrsim 1$. Also, the minimum amplitude of the forcing to restore the high--amplitude oscillations, for almost all the frequency values in the interval,
is $ F\approx0.1 $.

Although we are interested in small values of the forcing parameter $F$, we show in figures~\ref{fig:6}c-d an expansion of figure~\ref{fig:6}b. This gives us a better understanding of the
amplitude basins, by showing them for higher values of the forcing $F$ and for a more complete range of frequencies $\Omega$. In fact in figure~\ref{fig:6}c the value of the forcing $F$ goes from $0$ to $1$ and $\Omega$ goes from $0$ to $\pi$. In figure~\ref{fig:6}d the value of the forcing $F$ goes from $0$ to $10$ and $\Omega$ goes from $0$ to $2\pi$. In all the figures it is possible to appreciate that the effect of the forcing is either none or it suddenly makes the amplitude jump into a sustained oscillation between the two wells of the potential. If we compare the figures and the gradient bar on the right side of each of them it is possible to check the previous affirmation.

Finally, in figure~\ref{fig:6}d it can be seen that the amplitude basins for higher values of $F$ become more complex and that the role played by the frequency becomes really complicated. Also, for
$F\gtrsim2$ the higher amplitude for the oscillations appears around a different value of the forcing frequency, i.e., $\Omega=2$, the same behavior could be seen for $\mu=0.03$, but the figure is not in the panel for lack of further information. It appears that the parameter space can be divided in two areas, for small forcing values on the bottom of the figure, the system is more sensitive to the $\Omega=1$ frequency, while for $F>2$ the system frequency sensitivity shifts towards $\Omega=2$. So that, it makes that value of the forcing amplitude a critical value that divide two different behaviors of the system, at least for the Region IV, so we call it $F_{\mathrm{c}}^{\mathrm{IV}} \approx 2$.

\subsection{Forcing amplitude to recover the interwell oscillations in Region IV}

\begin{figure}[htp]
    \centering
    \includegraphics[width=13.5cm,clip=true]{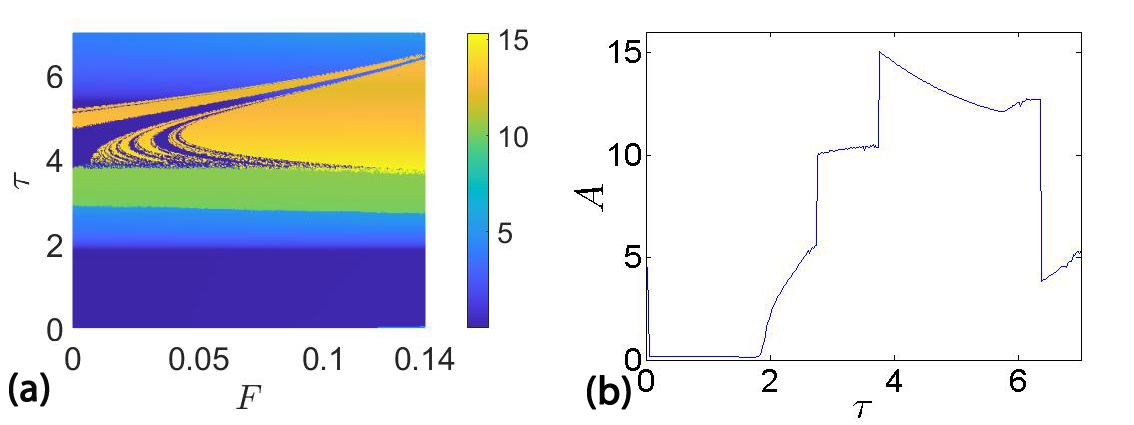}
\caption{(a) The amplitude basins in the $(F, \tau)$ parameter space for $\mu=0.04$, $\Omega=1$, $\tau\in[0.01,7]$ and $F\in[0,0.14]$. Panel (b) shows a vertical slice along the $\tau$ values of panel (a) for $F=0.14$, to better visualize the reconstructed amplitudes and the similarities with figure~\ref{fig:1}a. }
    \label{fig:7}
\end{figure}

Finally, it is relevant to study the effects of the forcing amplitude in the same way we did for the dissipation. Therefore, we decided to plot, in figure~\ref{fig:7}a, the $(F,\tau)$ parameter space, for $\mu=0.04$ and $\Omega=\Omega_{\mathrm{r}}$, and compare it with the $(\mu, \tau)$ plot of figure~\ref{fig:3}a. The value of the damping term has been fixed in order to study the possibility to restore the
interwell oscillations for a case where the oscillations are completely damped.

It can be seen in figure~\ref{fig:7}a how the high-amplitude oscillations (yellow region) are  restored for increasing values of $F$.
%
%
%
Interestingly, the figure shows a mirror symmetry with respect to figure~\ref{fig:3}a. The system's dynamics can be checked in figure~\ref{fig:7}b, where we represent a slice of figure~\ref{fig:7}a, for $F=0.14$. It is possible to see that the figure is similar to figure~\ref{fig:1}a, that displays the evolution of the amplitude versus $\tau$ for the case $\mu=0$. In fact, in both figures~\ref{fig:7}a-b, it is possible to appreciate that all the regions that remained unaffected by the damping, are also not affected by the forcing. On the other hand, the system's behavior in Region IV is the same as with $\mu=0$, once the forcing amplitude reaches values close to $F=0.14$.

In the light of these results, we can confidently say that the damping-induced unpredictability has disappeared and, finally, the interwell oscillations have been restored by increasing a small amount the forcing parameter $F$. Also, in Region IV, the no-damping situation is fully recovered due to the interaction of the time delay and the forcing, i.e., the delay-induced resonance phenomenon.

\begin{figure}[htp]
    \centering
    \includegraphics[width=13.5cm,clip=true]{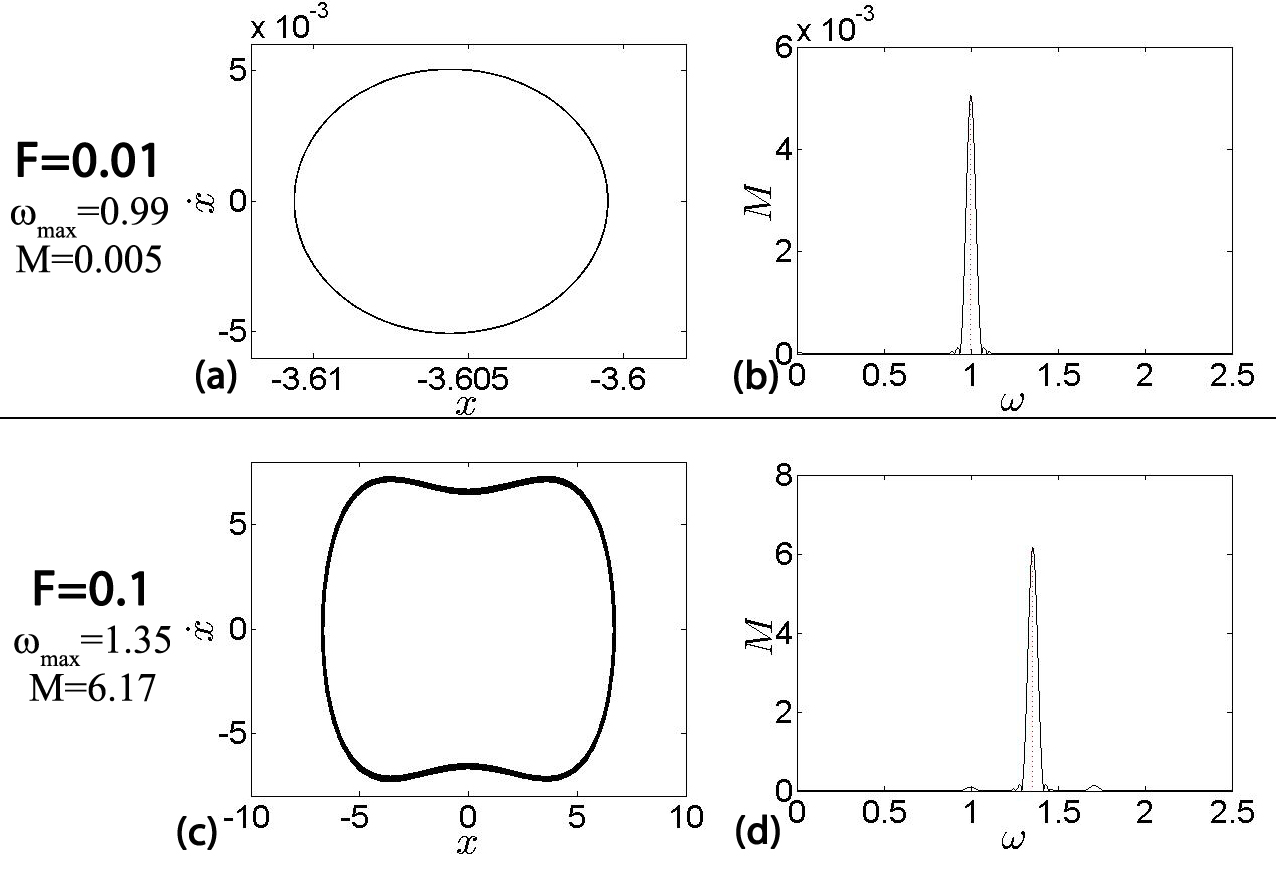}
\caption{Representation, from left to right, of the phase space orbits and the FFT of two trajectories. On panels (a) and (b) the parameters are set to fall before the beginning of the yellow basin of figure~\ref{fig:7}a, while in the others to fall inside it. In particular for both trajectories we set $\tau=4.5$, $\Omega=1$ and $\mu=0.04$, but for the upper one $F=0.01$ and for the lower one $F=0.1$. From the legend, note that in (b) $\omega_{\mathrm{max}}$ (marked in the figure as a red dotted line) matches the forcing frequency $\Omega=1$, while this does not occur in (d).}
    \label{fig:8}
\end{figure}

In figure~\ref{fig:8}, we show, from left to right, the phase space orbits and the FFT of two trajectories with different $F$ values, for a better understanding of actual result.
%
%
%
In the first case (figures~\ref{fig:8}a and b), for the chosen $F$  value, the parameters correspond to a region before the beginning of the yellow basin, while in the second one (figures \ref{fig:8}c and d) the parameters correspond to the yellow basin. It can be seen that in figure~\ref{fig:8}b the oscillation frequency matches the forcing frequency, while in figure~\ref{fig:8}d, when the delay-induced resonance is triggered, the oscillations frequency is different. Actually, the frequency of the sustained interwell oscillations is $\omega=1.35$ which is in agreement with the frequency of the delay-induced oscillations of the Region IV ($F=\mu=0$),  as we can check in figure~\ref{fig:2}h and its legend.

\begin{figure}[htp]
    \centering
    \includegraphics[width=13.5cm,clip=true]{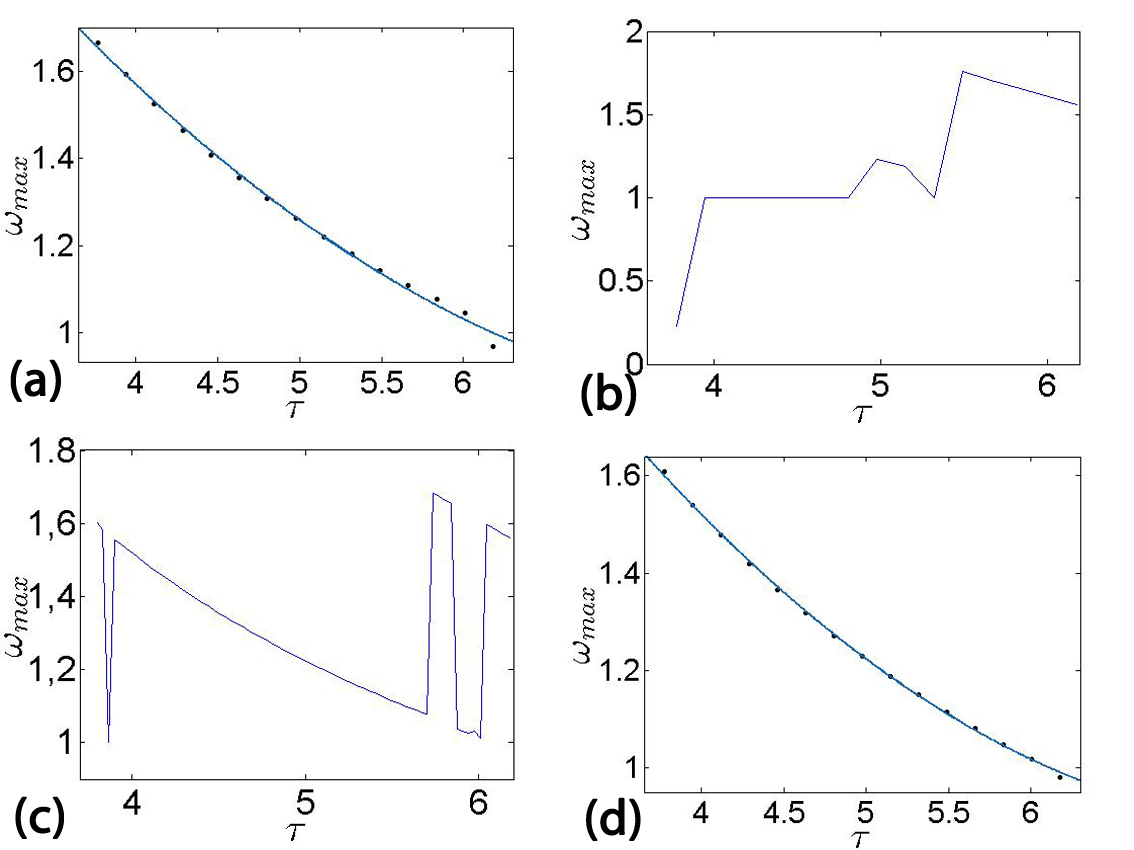}
\caption{The variation of $\omega_{\mathrm{max}}$ (the frequency of the delay-induced oscillations) with the time delay $\tau$ for (a) $F=\mu=0$, (b) $F=0.01$, (c) $F=0.1$ and (d) $F=0.14$ with  $\mu=0.04$ and $\Omega=1$. In (b) and (d) the solid circles are the numerically computed $\omega_{\mathrm{max}}$ while the continuous cuve is the polynomial fit (equation~\ref{eq:9}). }
    \label{fig:9}
\end{figure}

We depict, in figure~\ref{fig:9}, the Region IV oscillation frequencies, $\omega_{\mathrm{max}}$ calculated with the FFT, for different orbits changing the values of the parameters $\mu$ and $F$.
%
%
%
In particular, in figure~\ref{fig:9}a we show the frequencies of the delay-induced oscillations setting $F=\mu=0$. Then, in figures.~\ref{fig:9}b-d we show the oscillation frequencies for $\mu=0.04$ and $F=0.01$, $F=0.1$ and $F=0.14$, respectively. We can appreciate that as the forcing amplitude grows, the oscillation frequencies are restored to the no-damping case. In fact, in figure~\ref{fig:9}b, the frequencies are nothing similar to the non-damping case. On the other hand, in figure~\ref{fig:9}c, the frequencies curve is almost restored, except for some boundary $\tau$ values, that are related with the zones of the region on the parameter space in which there is still some remains of fractalization. Finally, figure~\ref{fig:9}d shows that the oscillation frequencies return close to the values of no-damping case. To corroborate it, we have plotted the curve fits on figures~\ref{fig:9}a-d. The relation is polynomial and reads
\begin{equation}\label{eq:9}
  \omega_{\mathrm{max}}(\tau) = p_1 \tau^2 + p_2\tau + p_3.
\end{equation}
The coefficients of the two polynomials are $p_1=0.04$, $p_2=-0.7$ and $p_3=3.6$, with deviations of small order.

The point is that the case without dissipation is restored, not only regarding the oscillations amplitude, but also regarding the oscillations frequencies. This is the effect of the conjugate effect, Ref.~\cite{Julia}, that suggests that if the oscillator is driven by a small forcing, we can enhance those oscillations, by adding a delay term. In this context, the delay plays the role of the forcing in triggering the resonance, so that the final oscillation frequency matches the delay frequency $\omega$ instead of the forcing frequency $\Omega$, like in our case.

\section{Delay-induced resonance for the single-well oscillations}\label{sec:4}

In the previous section, we focused on the effect of the forcing as an element to restore the high-amplitude oscillations that the damping had eliminated. In other words, we focused on the range of $ \tau $ for which the yellow amplitude basin disappeared. Now, we consider the effect of the forcing in the remainder of the range of $ \tau $. So, we shift our attention to the single-well oscillations of Regions II and V and explore the forcing parameter values, $F$ and $\Omega$ that trigger the delay-induced resonance and generate interwell oscillations in that region.

\begin{figure}[htp]
    \centering
    \includegraphics[width=13.5cm,clip=true]{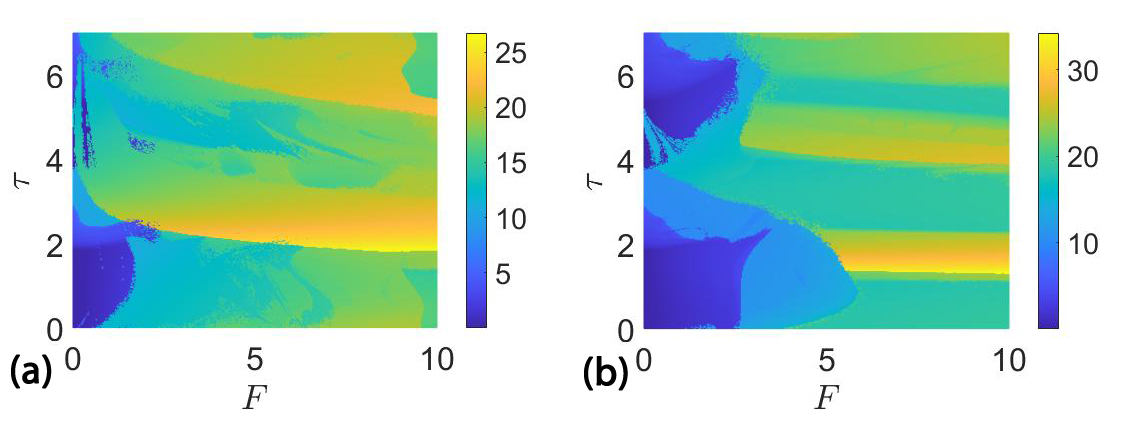}
\caption{The plot of the amplitude basins in the $(F,\tau)$ parameter space. Here $\mu=0.04$ and (a) $\Omega =1$  and (b) $\Omega =2$.}
    \label{fig:10}
\end{figure}

For values of $ \tau$ outside the Region IV, in figure~\ref{fig:7}a, the forcing does not have any effect. Thus, we need to increase the magnitude of the forcing in order to change the dynamics for the rest of the $ \tau $ values. Particularly, we are interested in the possibility of increasing the energy of the oscillations confined into one well, $\tau\in(2,3)$  and $\tau>6.18$ (light blue regions online in figure~\ref{fig:3}a), namely Region II and V, so that they become interwell oscillations. To that end, the amplitude basins in the $(F,\tau)$ parameter space for higher values of the $F$ are depicted in figure~\ref{fig:10}a-b for $ \Omega=1 $ and $ \Omega=2 $, respectively.
%
%
%
The first frequency choice has been taken to carry on the previous section analysis; the second one is related to figure~\ref{fig:6}d, where for a higher value of the forcing amplitude, the higher-amplitude oscillations pop up for precisely $\Omega=2$. In fact, in that figure, for $\tau=4.5$ and for $F>2$, the oscillations amplitude rises to $ A>20 $. Thus, in figure~\ref{fig:10}, we can see how the light blue amplitude basin reaches the interwell oscillations.

We conclude that it is possible to enhance the oscillations confined to one well of Regions II or V, so that they become interwell oscillations when the amplitude of the forcing increases. Also,
figures~\ref{fig:10}a-b show that the system's response for Region II is different depending on the frequency value. In fact, in both figures we can divide the Region II in 2 different areas: the
single-well oscillations area, on the left, and the interwell oscillations area, on the right. The two regimes  have a clear boundary in the two figures but different critical $F$ values. It is
possible to find those critical parameter values where the dark blue zone finish along with the single-well oscillations, at $\tau\approx2.2$ and for $F\approx2.4$ in figure~\ref{fig:10}a and
at $\tau\approx2.4$ and $F=3.5$ in figure~\ref{fig:10}b.

Moreover, in this last figure, we can divide the interwell oscillations area (Region II) in two subareas: one in the middle of the figure (the light blue online), where the system reaches the interwell oscillations and the other one on  the right (the yellow one online), in which the interwell oscillations rise at their maximum value, i.e., $ A \approx 30 $ or more.  It can be
appreciated, just as we commented before for figure~\ref{fig:6}d, that for a certain critical $F$ value the system sensitivity to the forcing frequency shifts from $\Omega=1$ to $\Omega=2$. For the
Region IV it was $F_{\mathrm{c}}^{\mathrm{IV}}\approx2$, for the Region II it is  $F_{\mathrm{c}}^{\mathrm{II}}\approx5$. So, we can say that those critical $F_c$ discriminate two different regimes in those two regions: small- and high-amplitude forcings. So that, the system is more sensitive to the frequency $\Omega'_{\mathrm{r}}=1$ before the critical value $F_\mathrm{c}$ and then becomes more sensitive to $\Omega''_\mathrm{r}=2$ for values of $F$ beyond
the critical one. Finally, also Region V arises to interwell oscillations in both figures but does not show the same behaviors as Region II. Therefore, the amplitude of the oscillations does not
change significantly from one plot to the other, although, for $\Omega=1$, the interwell oscillations start for a smaller value of the forcing amplitude.

\section{Conclusions}\label{sec:5}

We have analyzed the effect of the damping on the dynamics of the underdamped time-delayed Duffing oscillator. Firstly, we have shown that for small damping parameter values, high-amplitude oscillations (related with Region IV of time delay values), are damped while the rest of the dynamics is not affected. This effect of the damping produces a fractalization in the parameter space increasing the unpredictability of the system. We have demonstrated that this unpredictability can be reverted by a very small forcing amplitude with a specific value of the resonance frequency, $\Omega'_{\mathrm{r}}=1$, through the delay-induced resonance phenomenon. Note that this is a similar resonance effect as vibrational resonance where one of the external periodic forcings is substituted by the delay term. Moreover, not only the oscillations amplitude is restored, also the oscillations frequencies of the case without forcing and damping are restored, thanks to the effect of the conjugate phenomenon.

Then, we have found a critical value of the forcing amplitude, $F_{\mathrm{c}}^{\mathrm{IV}} = 2$ that switches the system sensitivity to the forcing frequency towards $\Omega=2$. So that, for $F>F_{\mathrm{c}}^{\mathrm{II}}$ the higher
oscillations amplitude are reached for $\Omega=2$, while for smaller values are reached for $\Omega=1$.

Finally, we proved that the same resonance phenomenon may be used to produce interwell oscillations for $\tau$ values inside the Regions II and V, for which the oscillations are bounded to one well, even without dissipation. The forcing amplitude, in these cases, needs to be of a bigger magnitude to induce the intrawell oscillations. Moreover, the forcing amplitude plays, again, a key role in the system sensitivity to the forcing frequency for Region II. In fact, we have found a threshold value of the forcing amplitude, namely $F_{\mathrm{c}}^{\mathrm{II}}\approx5$. For smaller values of the forcing amplitude, the system resonates for $\Omega=1$. While for values of $F$ higher than $F_{\mathrm{c}}^{\mathrm{II}}$, the resonance frequency is $\Omega=2$. Finally, we expect that this work can be useful for a better understanding of the delay-induced resonance phenomenon in presence of both dissipation and forcing. On the other hand, Region V reaches the interwell oscillations for a smaller value of $F$ when the forcing frequency is $\Omega=1$.

\section{Acknowledgment}

This work has been financially supported by the Spanish State Research Agency (AEI) and the European Regional Development Fund (ERDF, EU) under Projects No.~FIS2016-76883-P and No.~PID2019-105554GB-I00.


\end{document}